# Quantum key distribution with 1.25 Gbps clock synchronization

J. C. Bienfang, A. J. Gross, A. Mink, B. J. Hershman, A. Nakassis, X. Tang, R. Lu, D. H. Su, Charles W. Clark, Carl J. Williams

*National Institute of Standards and Technology, 100 Bureau Dr.,Gaithersburg, MD 20899*
*bienfang@nist.gov*

**E. W. Hagley, Jesse Wen**

*Acadia Optronics LLC, 13401 Valley Drive, Rockville, MD 20850*

**Abstract:** We have demonstrated the exchange of sifted quantum cryptographic key over a 730 meter free-space link at rates of up to 1.0 Mbps, two orders of magnitude faster than previously reported results. A classical channel at 1550 nm operates in parallel with a quantum channel at 845 nm. Clock recovery techniques on the classical channel at 1.25 Gbps enable quantum transmission at up to the clock rate. System performance is currently limited by the timing resolution of our silicon avalanche photodiode detectors. With improved detector resolution, our technique will yield another order of magnitude increase in performance, with existing technology.

---

## 1. Introduction

Cryptography requires secure and efficient key distribution. Quantum key distribution (QKD) [1] is provably secure, but it remains to be made efficient in real-world applications over distances and at bit rates consistent with the requirements of modern telecommunications. Previous work has demonstrated QKD over distances up to 150 km in fiber [2,3] and 23 km in free space [4,5], but the bit rates of these systems have been low. Typical single-photon link losses of QKD systems are about 30 dB, so the roughly 1 Mbps transmission rates of previous systems have resulted in sifted-key rates of the order of 1 kbps. Such rates are insufficient for network and telecommunications applications of the one-time-pad cipher, or for large numbers of multiple users. This paper reports a QKD system that has attained sifted-key rates of up to 1 Mbps over a 730 m free-space link, and identifies pathways for increasing the transmission rate by another order of magnitude.

A primary issue in all QKD systems is identifying the single transmitted photon in the presence of very high background, either from the sun in a free-space system or from single-photon detectors in a 1550 nm fiber system. Narrow temporal gating has been shown to be an effective technique for improving signal-to-noise levels. To date such gating has been applied in an asynchronous mode: each transmitted pulse, or burst of pulses, is preceded by a timing signal to which the receiver aligns the gate [4]. In contrast, our system uses standard 8B/10B encoding [6] and clock recovery at 1.25 Gbps to operate in a synchronous mode. This approach has enabled us to demonstrate continuous transmission rates that are significantly higher than any previously reported. High transmission rates serve both to enhance data encryption capabilities and to extend the distance over which a QKD system can operate. This article discusses the design and performance of our system, as well as demands placed on the photon source, data-handling algorithms, and timing resolution that are associated with GHz transmission rates.

## 2. Single-photon QKD

Quantum key distribution uses a quantum channel, comprised of a set of non-orthogonal bases, and an auxiliary classical channel to create a link over which two parties can develop a secret key. The security of the key developed by the sender (Alice) and receiver (Bob) is based on their ability to detect the measurements of an eavesdropper on the quantum channel. There are a variety of protocols that realize QKD by transmission of single photons. Our system is designed to implement the four-state BB84 protocol [7] with linear-polarization states, but in this initial demonstration we have implemented the B92 protocol [8,9]. This two-state protocol, though simpler to set up, generates sifted key at half the rate of the BB84 protocol.

To our knowledge, no single-photon-on-demand source operates at GHz repetition rates, so attaining such rates currently requires the use of an attenuated source. In our application, the quantum channel uses pulses attenuated to a mean photon number $\mu < 1$. The security issues associated with multiphoton pulses are a topic of current research [10,11], but a benchmark value for the mean photon number is $\mu = 0.1$. At this value about 9 % of the pulses contain a single photon, 1 % contain two or more photons, and the rest are empty. This results in a ten-fold reduction in maximum throughput due to the sources alone. High-speed single-photon-on-demand sources could provide an order of magnitude increase in the key generation rate.

## 3. Experimental setup

Figure 1 shows the layout of our experimental system. Alice and Bob are located inside two buildings separated by 730 m. A free-space 1550 nm optical link between the buildings provides four full-duplex wavelength-division multiplexed (WDM) channels at 1.25 Gbps. These channels comprise the classical channel for the QKD protocols: one we refer to as the primary classical channel (labeled $\lambda 1$ in Fig. 1), and is driven directly by custom data-handling PCI boards, while the others are used as dedicated Ethernet links between Alice and Bob. These systems have optical beacons and active tracking, and can operate over ranges of a few kilometers. This link operates continuously between the buildings.

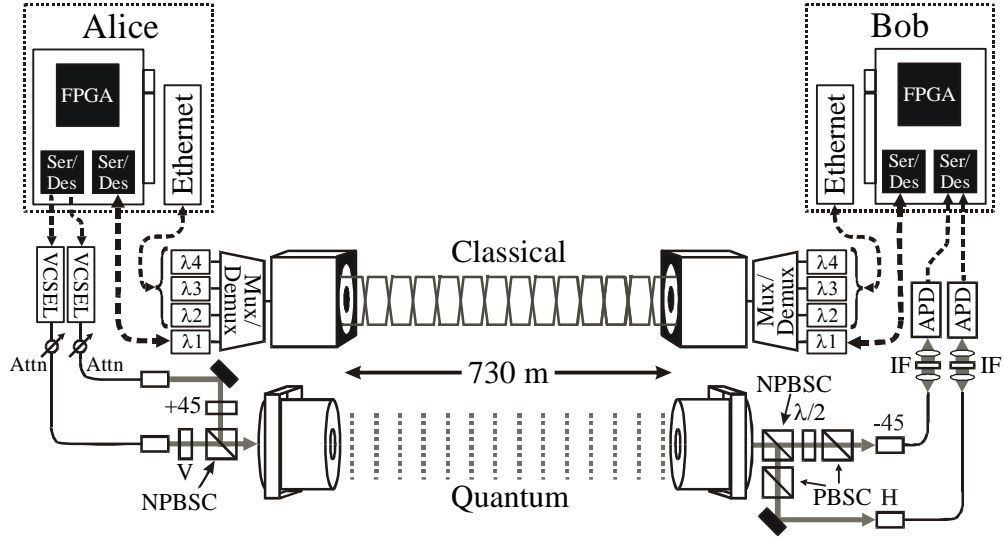

Fig. 1. The experimental setup. Alice and Bob are personal computers with custom data-handling PCI boards and gigabit Ethernet cards.

The quantum channel runs parallel to the classical channel and operates at 845 nm. The quantum channel sources are 10 GHz vertical-cavity surface-emitting lasers (VCSELs) and are driven by Alice's PCI board. The bias voltage on the VCSELs is set so that they produce 250 ps pulses with high extinction ratio, and we attenuate the pulses with variable fiber attenuators (Attn). The attenuated pulses are coupled, via single-mode fiber, to free-space optics mounted on the back of the transmit telescope where they are collimated, linearly polarized in either the vertical (V) or +45 degree direction, and then combined with a non-polarizing beam-splitting cube (NPBSC). The beam is then shaped to fill the entire 20.3 cm diameter output aperture of the Schmidt-Cassegrain telescope. The receive telescope is identical to the transmit telescope. A non-polarizing beam-splitting cube at the output of the receive telescope performs Bob's random choice of polarization state measurement: either horizontal (H), or –45 degrees. The measurements are made with polarizing beam-splitting cubes (PBSC) used in transmission, and we observe polarization extinction ratios in excess of 500:1 for both bases. The pulses transmitted by the PBSCs are coupled via 200-μm fiber to a detector box, where they pass through a 2 nm spectral filter (IF) and are then focused on to a silicon avalanche photodiode (APD).

The PCI boards at Alice and Bob each have a field-programmable gate array (FPGA), and two four-channel gigabit Ethernet serializers/deserializers (SerDes): one for the primary classical channel, and one for the quantum channel. The board clock rate is 125 MHz and the four SerDes operate on 10-bit words, resulting in 1.25 Gbps on each serial data channel. The quantum and classical channel SerDes on Alice's transmit board are synchronized by sharing the same clock. On the primary classical channel we use 8B/10B encoding to transmit a balanced 1.25 Gbps serial data stream to which the classical-channel SerDes at Bob can lock an internal phase-locked loop (PLL). The data received on the quantum channel is too sparse to synchronize Bob's quantum-channel SerDes. To synchronize this SerDes we mix the sparse quantum-channel data with the 8B/10B encoded signal from the primary classical channel with a logical exclusive-or (XOR) operation. The quantum-channel data is recovered in Bob's FPGA with another XOR operation. In this way we establish synchronization between Alice and Bob, and create a continuous series of 800 ps time bins used to gate the quantum channel.

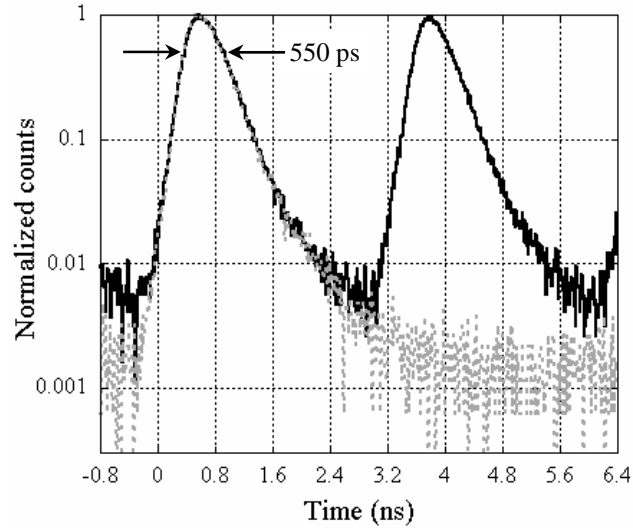

Fig. 2. Jitter in the APD from 250 ps optical pulses. The arrival times of the pulses, as reported by the APD, are histogrammed in 12.2 ps time bins. The black line shows a transmission rate of 312 MHz, the gray dashed line shows a transmission rate of 78 MHz. For both distributions the APD count rate is 100 kHz.

The FPGA on Alice's transmit board generates and stores the two 1.25 Gbps bit streams of random data necessary for the quantum channel basis and bit value in the BB84 protocol. Our random data is currently generated by a pseudo-random number generator on Alice's FPGA, but the board also has parallel inputs for an external random source. The data is organized in 2048-bit frames. The quantum channel sends the random data while the primary classical channel sends a synchronizing message, which includes a 32-bit frame number. At the receiver the rising edge of the detector signal is used to identify the time bin in which the photon arrived. For each detection event, the frame number, bit position, and a basis bit (for BB84) is returned to Alice over the primary classical channel, allowing her to sift her rapidly accumulating store of random data. To accommodate the inadvertent dropping of frames, Alice and Bob use the Ethernet channel to state which frames each has processed and passed on to memory. A frame is disregarded unless both have passed it on to memory. Sifting directly on our custom PCI boards provides a manageable data rate to the CPU for error correction and application-level data encryption. Both boards communicate with the CPU via a standard PCI interface using direct memory access. We are currently using a Linux operating system with custom drivers for the boards.

### 4. Performance and results

Although synchronization allows for quantum-channel transmission at the full 1.25 Gbps clock rate, we find that jitter in the APD limits us to lower transmission rates. Sending a series of attenuated 250 ps pulses at 312 MHz (every fourth bit) into an APD results in the distribution of APD detection events shown by the black line in Fig. 2. Also shown is a gray dashed line depicting the distribution when the transmission rate is only 78 MHz (every sixteenth bit). The full width at half maximum (FWHM) of these profiles are 550 ps. It was possible to reduce the FWHM of these profiles to 360 ps by reducing the spot size on the APD, but such spot sizes were unattainable with the fiber-coupled system. From the gray dashed line in Fig. 2 it can be seen that detection events from 250 ps optical pulses are distributed over roughly 3.5 ns. Such a distribution can generate errors in the sifted key at high transmission rates, when a detection event intended for a particular bin occurs in an adjacent time bin. Thus, even though both the custom PCI boards and the VCSELs are capable of operating at a 1.25 GHz repetition rate, we space our pulses by 4 bits to accommodate this jitter, resulting in a transmission rate of 312 MHz. In this mode the receiver masks each group of 4 bins by throwing away all the events that did not arrive in either the first or second bins, effectively operating with a 1.6 ns time gate. The FPGA disregards counts which occur on both APDs in the same gate. We find that this mask accepts 93 % of the intended detection events, and roughly 0.5 % from preceding pulses, corresponding to a 0.25 % error rate due to the quantum channel itself.

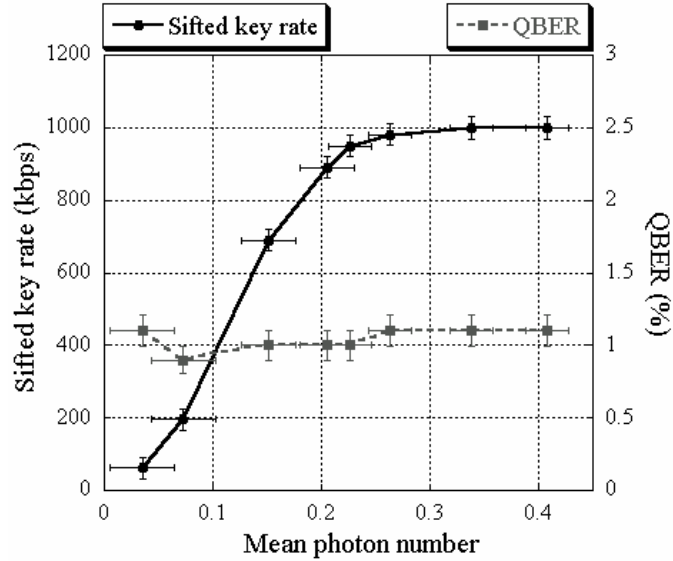

Fig. 3. Sifted key bit rate and quantum bit error rate vs. mean photon number.

The commercially available single-photon APDs used in this work are designed to maximize quantum efficiency, and timing resolution of 350 ps is typical for these devices when illuminated with a small spot. Previous research has shown that their timing resolution is ultimately limited to greater than 150 ps by photogenerated-carrier drift in the relatively thick depletion region. Single-photon-APD timing resolution below 50 ps FWHM has been demonstrated in devices with thinner depletion regions, but at reduced quantum efficiency [12]. Such devices will benefit the system presented here provided the gains in transmission rate outweigh the losses in detection efficiency.

Excluding the NPBSC and polarization measurement, we observe roughly 5 dB of loss from the output of the transmit telescope to the output of the 200 μm core fiber at the receiver. It is worthwhile to note that the central obscuration in the Schmidt-Cassegrain transmit telescope causes significant diffraction of the beam, which is likely to contribute to these losses. The transmissivity of the interference filters and focusing lens at the APD is 0.48. In spite of the 2 nm spectral filter, the daytime background count rate of the APDs (due to solar photons) exceeds 2 MHz, whereas the nocturnal background rate is on the order of 1 kHz on each detector. Improvements in filtering currently underway are expected to reduce the daytime background rate to the 100 kHz level, thereby enabling daytime operation.

A monitor at the NPBSC on the quantum transmit telescope is used to set the mean photon number of the pulses at the output aperture of the telescope. The sifted-key bit rates we observe as a function of mean-photon number are shown in Fig. 3. The system runs continuously, and we calculate the sifted-key bit rate from the amount of key saved to memory during a given amount of time, typically 60 seconds. For each value of the mean-photon number we also show the quantum-bit error rate (QBER). The QBER is the percentage of sifted key bits, produced by 60 seconds of key generation, for which Alice and Bob do not have the same value. At a mean photon number close to 0.15 we observe 690 kbps of sifted key at an error rate of 1.0 %. This compares well with the roughly 900 kbps one would expect for the B92 protocol, given the 312 MHz transmission rate, the losses stated above, and a detector quantum efficiency of 0.5. It can be seen from Fig. 3 that the bit rate increases with mean photon number up to 1.0 Mbps. Bit rates greater than 900 kbps put serious demands on the software, and we observe the system dropping a significant number of frames. This reduces the throughput of the system and causes the sifted-key rate to level off at 1.0 Mbps as the mean photon number increases above 0.2. Improvements in transfer across the PCI bus and the handling of interrupts are currently being implemented, and we expect this to increase the capacity of the system. The uncertainty in the mean-photon number accounts for observed variations in the VCSEL output amplitude.

Figure 3 shows that the QBER remains fairly constant at 1.1% as the key rate increases above 200 kbps. This fixed percentage suggests that at these key rates the majority of the errors are being generated by the quantum channel itself, rather than by background counts. It is worthwhile to note, however, that this error rate is higher than the 0.25 % we expect based on the APD jitter measurements shown in Fig. 2. Previous reports have shown that error correction techniques can be a bottleneck for QKD systems [4], potentially negating gains in throughput achieved on the quantum channel. Current research suggests that for a high-speed system it may be more productive to abandon the doctrine of bit preservation in favor of more expeditious reconciliation [13].

## 5. Conclusion

By leveraging high-speed optoelectronics technology, we have demonstrated transmission of quantum cryptographic key at rates of 1.0 Mbps, roughly two orders of magnitude faster than reported previously. This approach links the gate time of the quantum channel to the transmission rate, suggesting that it may be possible to realize greater throughput while reducing exposure to background photons. However, such an increase relies on improved detector resolution. Jitter in the silicon APDs has been shown to be a key limiter of performance, and imminent improvements in single-photon detectors will translate directly into higher transmission rates without increasing the error rate. This approach will not only enhance the throughput and reach of QKD systems, but will continue to yield dividends when the clock rate is increased.

**Acknowledgements**

This work was supported by the Defense Advanced Research Projects Agency under the DARPA QuIST program. We thank Eric Korevaar for helpful advice on free-space optics, and Richard Hughes and Christian Kurtsiefer for useful discussions.